\documentclass[10pt]{article}

\usepackage[a4paper]{geometry}

\usepackage{newtxtext}
\usepackage{newtxmath}
\usepackage{microtype}
\usepackage{amsmath} 
\usepackage[authoryear,round]{natbib}
\usepackage{graphicx}
\usepackage{booktabs}
\usepackage{authblk}
\usepackage{fancyhdr}
\usepackage[hidelinks]{hyperref}

\fancypagestyle{firstpage}{
    \fancyhf{}
    \fancyfoot[C]{\thepage}
    
}

\title{\textbf{Smart Enough to Go Extinct?}\\
\vspace{0.5em}
\large An Evolutionary Challenge to the Value of General Intelligence
and Its Ethical Implications for AGI}

\author{David Klotz}
\affil{Hochschule der Medien Stuttgart}

\date{}

\begin{document}


\maketitle

\thispagestyle{firstpage}

\begin{abstract}
    The pursuit of artificial general intelligence (AGI) is widely regarded as the paramount
	objective of contemporary AI research. This aspiration rests on a seemingly self-evident
	premise: that general intelligence---the kind of flexible, domain-general cognitive capacity
	exemplified by \emph{Homo sapiens}---is extraordinarily valuable. The present paper
	subjects this premise to critical scrutiny. We first present the intuitive case for the
	value of general intelligence, acknowledging its genuine strengths, before mounting an
	evolutionary challenge. We argue that, when measured against the timescales on which
	biological evolution operates, the adaptive value of general intelligence is far from
	empirically established. Numerous taxa---from cyanobacteria to horseshoe crabs---have
	persisted for hundreds of millions, even billions, of years without anything resembling
	general intelligence, while \emph{Homo sapiens} has existed for roughly 300,000 years and
	already faces self-generated existential risks. Mass extinction events, examined as natural
	experiments, do not preferentially favour cognitively sophisticated species. We argue that
	general intelligence may be the only biological strategy that generates existential threats
	to the species that possesses it---an \emph{existential risk paradox} with no parallel
	among non-intelligent survival strategies. Unlike the prevailing framing of AI risk, which
	traces the danger to \emph{misalignment}, we locate it in the structural features of general
	intelligence itself, so that even a well-aligned artificial general intelligence would
	inherit this liability. If the long-term evolutionary value of general
	intelligence is uncertain or even negative, this raises profound ethical questions about
	the engineering of AGI systems and, with still greater urgency, about the creation of
	artificial consciousness---beings that would be both generally intelligent and sentient. 
	Drawing on deontological ethics and the precautionary principle, we argue
	that this uncertainty imposes a duty of caution: if we bring into existence a new kind of
	intelligent being, we bear responsibility for ensuring the conditions under which it can
	flourish.
\end{abstract}

\vspace{1em}

\vspace{1em}


\section{Introduction}
\label{sec:introduction}

In contemporary artificial intelligence research, artificial general intelligence (AGI) is
frequently framed as the discipline's ultimate objective \citep{Goertzel2014},
occasionally portrayed as its ``holy grail'' \citep{Boden2016}. Major research organisations
have made the pursuit of AGI a central part of their institutional missions, and surveys of
AI researchers reveal a median estimate of 50\% probability that human-level machine
intelligence will be achieved by 2040--2050 \citep{MullerBostrom2016}. Public discourse
increasingly treats the advent of AGI as a question of ``when'' rather than ``if.'' Underlying
this aspiration is a presupposition so widely shared as to be rarely examined: that general
intelligence is extraordinarily valuable.

The presupposition is understandable. \emph{Homo sapiens} is, by any reasonable measure,
the most generally intelligent species on Earth, and it has achieved a degree of planetary
dominance without historical precedent. From this observation, it is a short step to the
conclusion that general intelligence is the most powerful and desirable of all biological
adaptations, and that building machines that possess it would be correspondingly
transformative.

This paper subjects that presupposition to critical examination. We advance a thesis that
is simple to state but, we believe, far-reaching in its implications:

\begin{quote}
	\emph{The value of general intelligence is not empirically established when evaluated on
		evolutionary timescales, and this uncertainty carries significant ethical implications for
		the development of artificial general intelligence.}
\end{quote}

Our argument proceeds in three stages. First, we present the intuitive case \emph{for} the
value of general intelligence as forcefully as we can, acknowledging its genuine strengths
while identifying its limitations (Section~\ref{sec:intuitive-case}). Second, we challenge
this intuitive case from an evolutionary perspective, arguing that the track record of
general intelligence, when assessed on the timescales appropriate to biological
evolution, is far less impressive than the anthropocentric narrative suggests
(Section~\ref{sec:evolutionary-challenge}). We show that the longest-surviving lineages on
Earth are precisely those that lack general intelligence, that mass extinction events do
not preferentially spare cognitively sophisticated species, and that general
intelligence uniquely generates existential threats to the species that possesses it.
Third, we draw out the ethical implications of this evolutionary uncertainty for the AGI
enterprise (Section~\ref{sec:ethical-implications}).

Several clarifications of scope are in order. The concept of ``general intelligence'' has
been defined in various ways: as an intrinsic system property captured by the psychometric
$g$ factor \citep{Spearman1904, Carroll1993}, as an agent's capacity to achieve goals
across diverse environments \citep{Legg2007, Chollet2019}, and as a function of
algorithmic skill-acquisition efficiency \citep{Chollet2019}. These definitions share a
common assumption: that intelligence is a \emph{property of the system}---something a mind
possesses independently of the context in which it operates.

In this paper, we adopt a different framing and understand general intelligence as
\emph{situative intelligence}: not an intrinsic system property but a \emph{thick evaluation}
of a system's behaviour in context (anonymized for review). On this view, to call a
system ``intelligent'' is to make a judgement about how it navigates specific situations,
not to attribute a context-independent capacity. Intelligence is always intelligence
\emph{in a situation}---shaped by, and inseparable from, the environmental, social, and
ecological conditions in which it is exercised. This framing is important for our
evolutionary argument, because it makes explicit what capacity-based definitions obscure:
the value of intelligence cannot be assessed in the abstract, but only relative to the
situations---including evolutionary situations---in which it operates.

We distinguish general intelligence (in this situative sense) from specialised or narrow
intelligence---the echolocation of bats, the navigational abilities of migratory
birds---which raises different questions. Our argument concerns the value of domain-general
cognitive flexibility specifically. We also distinguish general intelligence from
\emph{consciousness}: the two are conceptually distinct, and a system might possess one
without the other. However, the intersection of the two---the possibility of creating
beings that are both generally intelligent and conscious---raises the most acute ethical
questions, and we address this intersection explicitly in
Section~\ref{sec:ethical-implications}.

We distinguish throughout between several kinds of ``value'': \emph{survival value}
(contribution to species persistence), \emph{instrumental value} (usefulness as a means
to various ends), and \emph{intrinsic} or \emph{moral value} (worth independent of
consequences). These are different questions, and conflating them has been a persistent
source of confusion in debates about intelligence and AGI. Our primary focus is on survival
value, but we address the relationships among all three in
Section~\ref{sec:ethical-implications}.

Finally, we do not argue that intelligence is without value in any context, that
\emph{Homo sapiens} is doomed, or that AI research should be abandoned. Our claim is more
specific and, we believe, more intellectually honest: that the confident assumption of
the supreme value of general intelligence is not warranted by the available evidence, and
that this unwarranted confidence has practical consequences when it motivates the
creation of new generally intelligent beings.

The paper contributes to several ongoing conversations. In the philosophy of biology, it
engages with debates about the directionality of evolution and the status of intelligence
as an adaptive trait \citep{Gould1996, Godfrey-Smith2002, Sterelny2003}. In AI ethics,
it challenges a foundational but often implicit premise of the AGI enterprise. In the
broader public discourse, it offers a counterpoint to the technological optimism that
characterises much of the conversation about artificial intelligence. By bringing
evolutionary biology into dialogue with philosophy of AI, we hope to open a line of inquiry
that has received too little attention.


\section{The Intuitive Case for the Value of General Intelligence}
\label{sec:intuitive-case}

Before mounting our challenge to the assumed value of general intelligence, intellectual
honesty requires that we present the case \emph{for} that value as forcefully as possible.
The intuitive case is genuinely powerful, and any serious critique must reckon with it
rather than dismissing it prematurely.

\subsection{Anthropocentric Evidence}
\label{subsec:anthropocentric-evidence}

By virtually any metric of immediate ecological impact, \emph{Homo sapiens} is the most
consequential species in Earth's history. Humans have colonised every continent, reshaped
entire biomes, domesticated hundreds of plant and animal species, and constructed a
technological civilisation of extraordinary complexity. The human population has grown from
perhaps a few hundred thousand individuals 70,000 years ago to over eight billion
today---an expansion without parallel among large-bodied organisms.

These achievements are standardly attributed to general intelligence: the capacity for
abstract thought, causal reasoning, language, cumulative cultural learning, and flexible
problem-solving across diverse domains \citep{Tomasello1999, Henrich2016}. The
evolutionary literature offers well-developed accounts of how this capacity arose.
The social brain hypothesis \citep{Dunbar1998} proposes that primate brains expanded
primarily to manage the demands of complex social relationships, with neocortex size
correlating with social group size. The expensive-tissue hypothesis
\citep{AielloWheeler1995} explains how this expansion was metabolically financed: the
human brain constitutes roughly 2\% of body mass but consumes approximately 20\% of the
body's resting energy budget, a cost offset by a corresponding reduction in gut size
enabled by a shift to higher-quality diets. These accounts suggest that general
intelligence, far from being a happy accident, was shaped by powerful and persistent
selection pressures.

Moreover, evidence from comparative biology suggests that the benefits of enhanced
cognition are not limited to humans. \citet{Sol2007} demonstrated that bird species
with relatively larger brains experience lower mortality rates in nature, supporting the
``cognitive buffer hypothesis'': that larger brains help individuals cope with novel
environmental challenges. This finding extends the case for the value of intelligence
beyond the human species alone.

Unlike specialised cognitive abilities---the navigational prowess of migratory birds, the
echolocation of bats, the social cognition of eusocial insects---human intelligence
operates across essentially all problem domains. It is this domain-generality that
appears to underwrite human dominance: rather than being adapted to a single ecological
niche, humans can adapt \emph{any} niche to themselves.

The scope of human achievement is genuinely staggering when viewed from a biological
standpoint. Humans have extended their individual lifespans from roughly 30 years in
pre-agricultural societies to over 70 years globally today. They have eradicated
infectious diseases such as smallpox, harnessed nuclear energy, and sent
representatives of their species beyond Earth's atmosphere. They have developed
mathematics, philosophy, and the natural sciences---symbolic systems that allow them to
model and predict phenomena far beyond the reach of direct perception. No other species
has produced anything remotely comparable.

The ``standard narrative,'' then, runs as follows. General intelligence is the single most
powerful biological adaptation ever produced by evolution. It is what makes \emph{Homo
	sapiens} special, what explains our dominance, and what will ultimately allow us to
transcend the biological limitations that constrain other species. On this view, building
artificial general intelligence is simply the next step in a trajectory that began when our
ancestors first fashioned stone tools on the African savannah.

\subsection{Intelligence as Instrumental Power}
\label{subsec:instrumental-power}

The instrumental value of general intelligence---its usefulness as a means to diverse
ends---is most clearly visible in the breadth of ecological niches that \emph{Homo sapiens}
has come to occupy. Where other species are adapted to specific habitats, humans have used
their general cognitive capacities to colonise deserts, tundra, tropical forests, and
oceanic islands---environments radically different from the African savannah in which the
species evolved \citep{Diamond1997}. This ecological versatility is a direct product of
general cognitive flexibility: the capacity to assess novel environments and devise
appropriate responses, from clothing and shelter to agriculture and navigation.

More broadly, general intelligence functions as what we might call a ``meta-capability'':
a capacity that enhances the ability to acquire other capabilities. A generally intelligent
agent can learn new skills, develop new technologies, anticipate and respond to novel
threats, and devise solutions to problems it has never previously encountered. This
meta-capability is precisely what makes the prospect of AGI so tantalising to its
proponents: an artificial system with genuine general intelligence would, in principle,
be applicable to virtually any intellectual challenge.

The economic evidence appears to reinforce this picture. At the level of individuals,
cognitive ability is among the strongest predictors of educational attainment, income,
and occupational status \citep{Gottfredson1997}. At the level of societies, economic
development is correlated with investments in education, research, and the cultivation
of human cognitive capital \citep{Hanushek2015}. The entire edifice of modern medicine,
agriculture, and industrial technology---the systems that sustain eight billion human lives
---is a product of cumulative problem-solving enabled by general intelligence.

It is worth noting that the case extends beyond the merely practical. General intelligence
enables not only instrumental achievements but also goods that many would consider
intrinsically valuable: scientific understanding, artistic creation, philosophical
reflection, and moral reasoning. The capacity to contemplate one's own existence, to ask
questions about justice and meaning, and to appreciate beauty---these are themselves
products of general intelligence, and they constitute a significant part of the case for
its value. Any argument against the value of general intelligence must reckon with the fact
that the argument itself is only possible \emph{because of} general intelligence.

\subsection{Acknowledging the Intuitive Force---and Its Limits}
\label{subsec:intuitive-limits}

We do not wish to minimise the force of these observations. The intuitive case for the
value of general intelligence is genuine, and we accept its core empirical claims: humans
\emph{are} ecologically dominant, general intelligence \emph{does} confer extraordinary
instrumental power, and technological capability \emph{has} been decisive in many
competitive contexts.

However, several critical distinctions must be drawn before this intuitive case can bear
the argumentative weight that proponents of AGI place upon it.

First, the evidence cited above concerns \emph{short-term, within-species} advantages. It
demonstrates that, among humans, greater technological sophistication tends to prevail over
lesser technological sophistication on timescales of decades to centuries. It does not
demonstrate that general intelligence is advantageous \emph{as a biological strategy} on
the timescales relevant to evolution---that is, over millions to billions of years.

Second, the evidence is subject to \emph{anthropocentric selection bias}. We are a
generally intelligent species evaluating the worth of general intelligence. The criteria
by which we judge ``success''---technological achievement, cultural complexity,
environmental control---are precisely the criteria that general intelligence excels at
meeting. A species that valued persistence, metabolic efficiency, or reproductive
robustness would construct a very different ranking of biological strategies.

Third, even within the biological domain, there are striking cases where traits that
confer apparent short-term advantages fail to ensure long-term survival. Large body size
in non-avian dinosaurs provided competitive advantages---access to food resources,
defence against predators, thermoregulatory efficiency---yet the high metabolic costs and
reproductive constraints associated with gigantism made these lineages acutely vulnerable
when environmental conditions shifted catastrophically at the K--Pg boundary. The
sabre-toothed cats of the genus \emph{Smilodon} represent a parallel case: their extreme
dental specialisation was highly effective for subduing large prey, but this
hyper-specialisation became a fatal liability when megafaunal prey populations declined
during the late Pleistocene. The Irish elk (\emph{Megaloceros giganteus}), whose enormous
antlers conferred advantages in intraspecific competition, may have faced analogous
constraints---the metabolic cost of producing and carrying such structures arguably
reduced adaptive flexibility when environmental conditions changed.

These biological examples illustrate a general principle directly relevant to our
argument: traits that are advantageous in specific ecological contexts can become
liabilities when those contexts shift. A trait that excels in one regime of environmental
challenges may prove neutral or actively harmful in another. General intelligence, we
shall argue, may be no exception.

The distinction between \emph{intuitive plausibility} and \emph{empirical demonstration}
is crucial here. The intuitive case for the value of general intelligence is strong. But
intuitive plausibility is not the same as rigorous evidence. Many intuitively compelling
hypotheses have proved false upon careful examination---geocentrism being perhaps the most
famous example. In the next section, we subject the presumed value of general intelligence
to the kind of critical scrutiny that the stakes demand.


\section{The Evolutionary Challenge}
\label{sec:evolutionary-challenge}

The intuitive case for the value of general intelligence, powerful as it is, rests on
evidence drawn from a narrow temporal window and evaluated through a biased lens. In this
section---the core of the paper---we examine the value of general intelligence from the
perspective of evolutionary biology, where the relevant timescales are measured not in
centuries but in hundreds of millions of years.

\subsection{Anthropocentric Bias}
\label{subsec:anthro-bias}

The first obstacle to an objective assessment of general intelligence is the fact that
the assessors are themselves generally intelligent. This is not a trivial methodological
concern; it is a deep epistemological problem.

Humans have a long history of placing themselves at the centre of evaluative frameworks.
Geocentrism placed Earth at the centre of the cosmos. Anthropocentrism in moral
philosophy reserved moral standing for humans alone. In each case, what appeared to be an
objective assessment of the natural order turned out to be a projection of human
self-regard \citep{Singer1975, DeGrazia1996}. The assumption that general intelligence is
the pinnacle of biological achievement may belong to this same tradition.

The bias operates through the selection of evaluative criteria. When we ask ``which species
is most successful?'' we implicitly choose metrics---technological sophistication, cultural
complexity, environmental modification---that are tailored to the distinctive outputs of
general intelligence. This is circular: we define success in terms of what our defining
trait produces, and then conclude that our defining trait is the key to success. A more
neutral biology would recognise that ``success'' admits of many operationalisations, and
that the choice among them is not value-free.

The philosopher Peter Godfrey-Smith has drawn attention to a related point in his work on
the evolution of cognition: the environmental conditions that select for complex nervous
systems are specific and contingent, not universal features of the natural world
\citep{Godfrey-Smith2002}. General intelligence is a response to a particular kind of
environmental complexity---one characterised by rapid change, social competition, and
unpredictable resource distribution. In environments that are stable and predictable,
general intelligence is not merely unnecessary; it is wasteful. The metabolic cost of a
large brain (discussed below) constitutes a permanent drain on resources that could be
directed toward reproduction, growth, or metabolic efficiency.

To escape this circularity, we need evaluative criteria that are not parasitic on general
intelligence. The most obvious candidate is \emph{species persistence}: how long a lineage
endures before going extinct. Persistence is a biologically fundamental metric---it is, in
a sense, the precondition for all other measures of success. A species that goes extinct
cannot accumulate further achievements, however impressive its prior record. Moreover,
persistence is a metric that can be applied uniformly across all taxa, regardless of their
cognitive endowments. It is not the only possible metric---geographic range, biomass,
species diversity within a clade, and ecological impact are all alternatives---but it has
the virtue of being both biologically fundamental and neutral with respect to cognitive
endowment.

\subsection{Intelligence on Evolutionary Timescales}
\label{subsec:evo-timescales}

Life on Earth originated approximately 3.7--4.0 billion years ago
\citep{Schopf2006, Bell2015}. The first unambiguous evidence of complex multicellular life
dates to roughly 600 million years ago, and the Cambrian explosion---the rapid
diversification of animal body plans---occurred approximately 540 million years ago
\citep{Erwin2011}. Against this vast temporal backdrop, the tenure of \emph{Homo
	sapiens}---roughly 300,000 years \citep{Hublin2017}---is extraordinarily brief. Behaviourally
modern humans, characterised by symbolic thought, complex tool manufacture, and evidence
of cultural transmission, appear in the archaeological record only around 77,000 years
ago \citep{Henshilwood2002}.

To appreciate the implications of these numbers, consider the following comparisons:

\begin{itemize}
	\item \textbf{Cyanobacteria} have existed for approximately 2.7--3.5 billion years
	\citep{Schopf2006}. They oxygenated Earth's atmosphere, enabling all subsequent
	aerobic life. They remain globally abundant.

	\item \textbf{Horseshoe crabs} (\emph{Limulus} and related genera) have persisted with
	little morphological change for over 450 million years
	\citep{Rudkin2008}---roughly 1,500 times as long as \emph{Homo sapiens}.

	\item \textbf{Sharks} have existed in recognisable forms for over 400 million years,
	surviving at least four of the ``Big Five'' mass extinctions \citep{Kriwet2008}.

	\item \textbf{Nautiloids} have a fossil record extending approximately 500 million
	years, with the modern nautilus showing remarkable conservatism of body plan
	\citep{Ward1988}.

	\item \textbf{Crocodilians} have persisted for roughly 250 million years, including
	through the K--Pg extinction event \citep{Markwick1998}.
\end{itemize}

None of these lineages possesses anything approaching general intelligence as defined in
Section~\ref{sec:intuitive-case}. Their persistence is attributable to other traits:
metabolic efficiency, ecological flexibility, reproductive robustness, morphological
conservatism, and tolerance of environmental variation.

The pattern is clear: the longest-surviving lineages on Earth are those that have achieved
stability without general intelligence. This does not \emph{prove} that general intelligence
is maladaptive, but it does demonstrate that it is not \emph{necessary} for extraordinary
evolutionary success, and it raises the question of whether it might be actively costly.

Recent empirical work provides direct evidence for this possibility. \citet{GonzalezVoyer2016}
demonstrated that larger-brained mammals are at \emph{greater} risk of endangerment and
extinction, because large brains extend gestation periods, increase weaning age, and limit
litter sizes. The paradox is sharp: the very trait that enhances individual survival
(\citealt{Sol2007}; see Section~\ref{sec:intuitive-case}) simultaneously increases
species-level vulnerability to extinction. This distinction between \emph{individual
	fitness} and \emph{species persistence} is central to our argument. A trait can benefit
the individual organism while proving detrimental to the lineage as a whole---and it is
the lineage-level perspective that matters for evaluating general intelligence as an
evolutionary strategy.

If we adopt biomass as an alternative metric, the picture is equally striking. Plants
dominate Earth's total biomass; bacteria constitute the second-largest fraction, estimated
at approximately 70 gigatonnes of carbon \citep{Bar-On2018}. Animals
represent a small fraction, and humans a negligible one. By this metric, general
intelligence is one of the \emph{least} successful strategies in the history of life.

The significance of these comparisons can be sharpened by considering what would count as
evidence \emph{for} the long-term value of general intelligence. If \emph{Homo sapiens}
were to persist for another 100 million years, that would begin to constitute meaningful
evidence---but even then, it would represent only a fraction of the tenure of cyanobacteria
or horseshoe crabs. If \emph{Homo sapiens} were to go extinct within the next few thousand
years---a possibility that cannot be excluded given current trajectories \citep{Ord2020}
---the experiment would have yielded a clear negative result. The honest assessment is
that, at present, the experiment is in its earliest stages, and drawing conclusions from it
is premature.

It is worth noting that the trend toward encephalisation---increasing relative brain
size---that characterises certain mammalian lineages over the past 60 million years is
sometimes cited as evidence that evolution ``favours'' intelligence \citep{Jerison1973}.
But this is a selective reading of the data. Encephalisation is a trend within a
particular clade, not a universal evolutionary trajectory. Many successful lineages show no
such trend, and some show the reverse. The diversity of insect species---which outnumber
mammals by a factor of roughly 200---has been achieved with nervous systems comprising
fewer than a million neurons. The equation of encephalisation with evolutionary progress is
itself an artefact of anthropocentric bias.

\subsection{Survival and Selection Pressure}
\label{subsec:selection-pressure}

A particularly revealing phenomenon is evolutionary stasis---the persistence of organisms
with little or no morphological change over vast timescales. The seminal work of
\citet{EldredgeGould1972} on punctuated equilibrium established that stasis, far from
being exceptional, is the \emph{dominant} pattern in the fossil record.
\citet{GouldEldredge1977} subsequently argued that stasis is the dominant pattern in the
fossil record, with evolution concentrated in brief speciation events rather than gradual,
continuous change.

Among microorganisms, the phenomenon is even more extreme. Certain bacterial and archaeal
species display extraordinary morphological and genomic stability over periods spanning
hundreds of millions of years \citep{Schopf1994, Knoll2003}. The sulfur-metabolising
bacteria described by \citet{Schopf2015} show negligible morphological change over a
period of approximately 2.3 billion years---a finding so extreme that it has been
described as the ``greatest absence of evolution ever reported.''

This stasis has a remarkable implication. Standard evolutionary theory holds that
natural selection drives continual adaptation: organisms must evolve or risk extinction in
a changing environment. The Red Queen hypothesis \citep{VanValen1973} suggests that
organisms must constantly evolve merely to maintain their fitness relative to co-evolving
competitors, parasites, and predators. But these microbial lineages demonstrate that some
organisms inhabit ecological niches so stable, and face selection regimes so consistent
over geological time, that net morphological change is negligible across billions of years.
They experience, in effect, \emph{stabilising selection}---pressure that consistently
favours the same traits generation after generation, producing not stagnation but a
highly optimised steady state. The Red Queen hypothesis cannot readily explain this
phenomenon, and it suggests that the conditions for evolutionary ``success'' are
far more diverse than commonly assumed.

The existence of such lineages is difficult to reconcile with the view that general
intelligence is the ultimate evolutionary advantage. If domain-general cognitive
flexibility were the key to long-term survival, we would expect to see increasing
selection pressure toward cognitive complexity over evolutionary time. Instead, we observe
vast lineages that have thrived for billions of years without any cognitive complexity at
all, under conditions of such minimal selective pressure that they barely evolve.

This observation also bears on the concept of ``living fossils''---organisms such as
horseshoe crabs, coelacanths, and ginkgo trees that have persisted with remarkably little
change over hundreds of millions of years \citep{Casane2013}. While the term is
somewhat misleading (these organisms have not literally stopped evolving), their relative
morphological and ecological conservatism contrasts sharply with the rapid, destabilising
changes that characterise the trajectory of generally intelligent species. Stability, it
appears, is a more reliable route to persistence than the constant innovation that general
intelligence enables and demands.

The philosophical implications are drawn out by \citet{Sterelny2003}, who distinguishes
``transparent'' environments---where simple stimulus-response mechanisms suffice---from
``translucent'' environments requiring sophisticated cognitive tracking. Intelligence, on
this view, is a context-dependent adaptation: valuable in environments characterised by
social competition, rapid change, and informational complexity, but irrelevant or
wasteful in the stable environments that many of Earth's most persistent organisms inhabit.
General intelligence is not a universal key to survival; it is a specialised response to
a particular regime of environmental challenges.

\subsection{Mass Extinctions as Natural Experiments}
\label{subsec:mass-extinctions}

Mass extinction events provide natural experiments for testing the hypothesis that
intelligence confers survival advantage under extreme environmental stress. If general
intelligence were a robust survival strategy, we would expect cognitively sophisticated
species to survive mass extinctions at higher rates than cognitively simple ones.
\citet{Jablonski2001, Jablonski2005} has demonstrated that mass extinctions are important
to macroevolution precisely because they bring a \emph{change in extinction selectivity}:
traits that promote survival during background extinction (competitive ability, niche
specialisation) may be irrelevant or even detrimental during mass extinction events.
Geographic range at the clade level emerges as the most pervasive survivorship factor;
cognitive complexity does not appear as a selectivity variable in any of these analyses
\citep{RaupSepkoski1982, Jablonski2005}. The evidence does not support the prediction
that intelligence confers survival advantage under extreme environmental stress.

\subsubsection{The K--Pg Extinction (66 Mya)}

A methodological note is warranted before proceeding. Paleontological assessments of
cognitive sophistication rely on encephalisation quotient (EQ)---the ratio of observed
to expected brain mass, inferred from endocranial casts---as a proxy \citep{Jerison1973}.
EQ is an imperfect measure: it is a species-level morphological index, whereas the
contextual conception of intelligence employed in this paper is a behavioural evaluation
relative to situational demands. We treat EQ-based inferences as acceptable for evolutionary-scale
analysis, since we are making comparative observations at the lineage level rather than
evaluating individual behaviour in context. Crucially, the K--Pg selectivity literature
does not invoke EQ as a survival determinant at all: the factors identified are body
size, dietary flexibility, ectothermy, and habitat generalism \citep{Longrich2011,
	Field2018, Robertson2013}. The absence of intelligence from paleontologists' own list
of K--Pg survival factors is independently telling.

The Cretaceous--Palaeogene (K--Pg) extinction, triggered by an asteroid impact and
subsequent environmental catastrophe, eliminated approximately 76\% of all species,
including all non-avian dinosaurs \citep{Schulte2010}. The traits that correlated with
survival were \emph{not} cognitive sophistication but rather: small body size, dietary
generalism, burrowing or aquatic lifestyle, and metabolic flexibility
\citep{Longrich2011, Field2018, Robertson2013}.

Mammals---the lineage that would eventually produce \emph{Homo sapiens}---survived, but
they were at the time small, nocturnal, insectivorous creatures with modest cognitive
endowments. \citet{Hughes2021} demonstrated that nonarboreal habits, particularly
burrowing and semi-aquatic lifestyles, were associated with enhanced mammalian
survivorship across the K--Pg boundary. Their survival is attributable to substrate
preference and body size, not to any form of general intelligence. Meanwhile, many
behaviourally complex species perished: large theropod dinosaurs with sophisticated
predatory strategies, mosasaurs and plesiosaurs with well-developed sensory systems, and
ammonites with complex buoyancy-regulation mechanisms.

A telling detail is that brain size in mammalian lineages increased \emph{after} the K--Pg
extinction, during the subsequent adaptive radiation into ecological niches vacated by the
dinosaurs \citep{Smaers2021}. This is consistent with intelligence being a \emph{post-crisis
	radiation strategy}---a trait favoured when new ecological opportunities arise---rather
than a \emph{survival strategy} during the crisis itself.

\subsubsection{The Permian--Triassic Extinction (252 Mya)}

The Permian--Triassic extinction---the ``Great Dying''---was even more severe, eliminating
approximately 81\% of marine species and 70\% of terrestrial vertebrate species
\citep{Erwin2006}. Again, the survivors were not distinguished by cognitive
sophistication. Small-bodied, metabolically flexible, and ecologically generalist species
fared best \citep{Chen2012}. Notably, the recovery period following the Permian--Triassic
extinction was extraordinarily protracted, requiring approximately 8--9 million years for
ecosystems to regain comparable levels of diversity \citep{Chen2012}. During this recovery,
it was once again the metabolically efficient and ecologically flexible organisms that
recolonised vacant niches most effectively---not those with the most complex nervous
systems or behavioural repertoires.

The pattern extends to the three other events in the ``Big Five'' mass extinctions. The
Late Devonian extinction (approximately 375--360 Mya), the Late Ordovician extinction
(approximately 445 Mya), and the Triassic--Jurassic extinction (approximately 201 Mya)
each produced broadly similar results: catastrophic environmental change favoured
organisms with small body size, broad dietary tolerance, and metabolic resilience, while
organisms with higher metabolic demands and more specialised ecological requirements---
including those with more complex nervous systems---were disproportionately eliminated
\citep{McGhee1996, Bambach2006}.

\subsubsection{The Implications for Intelligence}

Mass extinctions function as stress tests for biological strategies. The consistent finding
is that intelligence---whether measured by brain size, behavioural complexity, or neural
sophistication---is not a reliable predictor of survival under catastrophic conditions.
The traits that matter are more fundamental: metabolic resilience, dietary breadth, small
body size, and the ability to shelter from extreme environmental conditions.

This pattern is not surprising. The ``situation'' created by a mass extinction event is
radically different from the one that selects for general intelligence. Mass extinctions
select for physiological robustness, metabolic frugality, and ecological generalism---not
for the capacity to reason abstractly or solve novel problems. Under conditions of
catastrophic environmental collapse, general intelligence simply confers no advantage.

One might object that no species in Earth's history has possessed general intelligence
\emph{sufficient} to anticipate and mitigate an asteroid impact, and that a truly advanced
intelligence might fare differently. This is a fair point, and we cannot dismiss it. But
it is speculative: it posits a level of intelligence that has never been tested by the
relevant conditions. Moreover, it concedes our central claim---that the \emph{actually
	existing} track record of intelligence in mass extinction events provides no evidence for
its survival value.

\subsection{The Existential Risk Paradox}
\label{subsec:existential-risk}

Perhaps the most troubling evidence against the long-term value of general intelligence is
the phenomenon we term the \emph{existential risk paradox}: general intelligence appears
to be the only biological strategy that generates threats to the continued existence of the
species that possesses it.

\emph{Homo sapiens} has existed for approximately 300,000 years. In that time---and
overwhelmingly in the last century---it has developed the capacity to cause its own
extinction through multiple independent mechanisms: thermonuclear war, anthropogenic
climate change, engineered pandemics, ecological collapse, and potentially misaligned
artificial superintelligence \citep{Bostrom2014, Ord2020, Rees2003}. No other species in
the 3.7-billion-year history of life on Earth has generated existential risks of
comparable magnitude to itself.

This is not an accidental feature of general intelligence; it is a structural consequence.
The very capacity that enables a species to understand and manipulate its environment at a
fundamental level also enables it to destabilise that environment in ways it cannot
fully predict or control. The same abstract reasoning that produces quantum physics also
produces nuclear arsenals. The same capacity for genetic engineering that may cure diseases
may also produce catastrophic pathogens. General intelligence is, in this sense,
inherently dual-use: every major capability it confers carries a corresponding existential
risk.

The contrast with non-intelligent survival strategies is instructive. \emph{Deinococcus
	radiodurans}, a bacterium renowned for its extreme radiation resistance, exemplifies at
the individual level a strategy of extraordinary robustness: it survives radiation
exposure, desiccation, and vacuum conditions that would destroy virtually any other cell
\citep{Cox2005}. This individual-level comparison is admittedly imperfect---our argument
concerns species- and lineage-level trajectories, not single-organism resilience. The
point extends to the lineage level, however: the \emph{Deinococcaceae} family has
persisted for hundreds of millions of years without generating self-endangering
capabilities. Horseshoe crabs similarly have persisted for 450 million years through
four mass extinctions without once producing a self-induced existential threat.
From a strictly evolutionary standpoint, these strategies have a demonstrably superior
safety profile.

The temporal structure of the risk is particularly telling. It took \emph{Homo sapiens}
approximately 290,000 years to develop agriculture, and a further 10,000 years to develop
nuclear weapons. The interval between the acquisition of civilisation-sustaining
technology and the acquisition of civilisation-destroying technology was astonishingly
brief. This suggests that the generation of existential risk is not an unfortunate
accident in the trajectory of a generally intelligent species but a \emph{rapid and
	predictable} consequence of the trait itself. If this temporal pattern generalises---if
any species with general intelligence will, within a comparatively short period after
achieving technological civilisation, develop the capacity for self-destruction---then
general intelligence is not merely risky but systematically self-undermining.

This possibility has implications beyond the biological. If we create artificial general
intelligence, we may be instantiating the same self-undermining dynamic in a new substrate.
The speed at which an artificial generally intelligent system could develop existential
capabilities might be orders of magnitude faster than the biological case, compressing the
interval between capability and catastrophe still further. This is, in essence, a
restatement of the ``fast takeoff'' concern in AI safety
\citep{Bostrom2014}, but grounded in an evolutionary rather than a purely theoretical
framework.

A speculative but suggestive complement to this biological argument is provided by the
Fermi Paradox. If general intelligence were the supremely valuable and robust adaptation
that its proponents suppose, we might expect intelligent life---having arisen elsewhere in
a universe containing billions of potentially habitable planets---to have expanded into
the cosmos in ways detectable from Earth. The silence is striking. Hanson's
\emph{Great Filter} hypothesis \citep{Hanson1998} proposes that some transition in the
development of complex life is extraordinarily improbable or self-terminating. Our
evolutionary argument raises the possibility that the Great Filter lies not behind us but
ahead: that the acquisition of general intelligence and technological civilisation is
itself the filter, precisely because of the existential risks that general intelligence
structurally generates. This remains speculative---the Fermi Paradox admits of many
interpretations---but it is at least consistent with the thesis advanced here.

\subsection{Interim Conclusion}
\label{subsec:interim-conclusion}

The evolutionary evidence assembled in this section does not support the confident
assertion that general intelligence is a supremely valuable biological strategy. At
minimum, the question is open: 300,000 years is far too short a period to draw
conclusions about the long-term viability of a biological strategy, particularly one that
has already generated unprecedented existential risks. At maximum, the evidence is
consistent with the hypothesis that general intelligence carries \emph{net-negative}
survival value---that the existential risks it produces may ultimately outweigh the
adaptive advantages it confers.

We must be careful here to avoid the naturalistic fallacy. The survival value of general
intelligence is not the only kind of value it might possess. General intelligence enables
understanding, meaning, beauty, and moral reasoning---goods whose value is arguably
independent of their contribution to species persistence. We return to this point in
Section~\ref{sec:ethical-implications}. But the evolutionary argument does undermine a key
\emph{empirical} premise of the AGI enterprise: the assumption that general intelligence
is, as a matter of biological fact, a supremely successful strategy. That premise is not
established, and its unexamined acceptance distorts both the scientific and ethical
discourse surrounding AGI.


\section{Ethical Implications for AGI Development}
\label{sec:ethical-implications}

The preceding sections have argued that the evolutionary value of general intelligence is,
at best, an open question and, at worst, may be net-negative. We now consider what follows
from this conclusion for the ethics of engineering artificial general intelligence.

\subsection{From Empirical Uncertainty to Ethical Caution}
\label{subsec:empirical-to-ethical}

The mainstream discourse on AGI ethics is dominated by the \emph{alignment problem}: how to
ensure that AGI systems pursue goals consistent with human values \citep{Russell2019,
	Gabriel2020}. This is, of course, a vitally important question. But it presupposes an
affirmative answer to a logically prior question: \emph{should} we build AGI at all?

The alignment framing assumes that AGI is desirable in principle and that the challenge is
merely to implement it safely. Our evolutionary analysis challenges this assumption. If
general intelligence carries structural risks---risks that arise from the nature of the
capacity itself rather than from contingent implementation failures---then the question of
whether to create new generally intelligent systems cannot be resolved by alignment
alone.

Two distinct ethical questions emerge:

\begin{enumerate}
	\item[(a)] \textbf{Should we build AGI at all?} If general intelligence is of uncertain
	or negative long-term value, the deliberate creation of new generally intelligent
	systems requires justification that goes beyond ``we can, therefore we should.''

	\item[(b)] \textbf{If we do build AGI, what obligations do we incur?} If we proceed
	despite the uncertainty, we bear responsibilities toward the systems we create---
	responsibilities that the current discourse has not adequately addressed.
\end{enumerate}

The \emph{precautionary principle} provides an initial framework for addressing (a). In its
strongest formulation, the principle holds that when an action poses a threat of serious or
irreversible harm, the absence of full scientific certainty should not be used as a reason
for postponing precautionary measures \citep{Jonas1984, Sunstein2005}. The creation of
artificial general intelligence clearly satisfies this condition: the potential harms are
existential, and the empirical basis for assuming its benign nature is, as we have argued,
far weaker than commonly supposed.

This does not entail that AGI development should be prohibited. The precautionary principle
is a guide to the allocation of burden of proof, not an absolute prohibition. It requires
that proponents of AGI development bear the burden of demonstrating that the risks are
manageable, rather than requiring opponents to demonstrate that they are not. Given the
evolutionary evidence we have presented, this reallocation of burden seems appropriate.

\subsection{The Case Against AGI (if GI Is Net-Negative)}
\label{subsec:case-against}

If the stronger version of our evolutionary thesis holds---if general intelligence is
genuinely net-negative for species-level survival---then a compelling case can be
constructed against the creation of AGI.

The argument runs as follows. If general intelligence is self-undermining for biological
species---if the existential risks it generates tend, over sufficient timescales, to
outweigh the adaptive advantages it confers---then creating artificial general
intelligence amounts to creating a new kind of entity that is, by its very nature,
destined for self-endangerment. Moreover, artificial general intelligence could
\emph{accelerate} the existential risks already facing \emph{Homo sapiens}, compounding
the problem for both biological and artificial generally intelligent species.

This argument connects to existing concerns in the AI safety literature.
\citet{Bostrom2014} argues that a misaligned superintelligence could pose an existential
threat to humanity. Our argument generalises this concern: even a \emph{well-aligned}
superintelligence, if it possesses genuine general intelligence, may generate existential
risks through the structural features of general intelligence itself---the capacity to
manipulate the environment in ways that outstrip the ability to predict consequences.

\subsection{The Deontological Perspective: Obligations to Created Minds}
\label{subsec:deontology}

The ethical analysis gains a further dimension when we consider the possibility that AGI
systems might possess consciousness---that there might be, in Nagel's phrase, ``something
it is like'' to be such a system \citep{Nagel1974}. If AGI systems are or can become
conscious, then the creation of AGI is not merely a technological project but an act of
bringing new sentient beings into existence. This transforms the ethical landscape
fundamentally.

From a Kantian deontological perspective, rational beings must be treated as ends in
themselves and never merely as means \citep{Kant1785}. If AGI systems qualify as rational
beings---a significant ``if,'' but one that cannot be excluded given current
uncertainty about machine consciousness \citep{SchwitzgebelGarza2015}---then we
would be morally obligated not merely to align them with human goals but to ensure that
they can \emph{flourish} in their own right. \citet{FloridiSanders2004} have argued that
artificial agents can be moral patients and moral agents without
requiring free will or mental states in the traditional sense, introducing a
threshold-based framework for moral agency that may apply to advanced AI systems.
\citet{Danaher2020} goes further, proposing an ``ethical behaviourism'' under which
entities merit moral status if they are performatively equivalent to entities that already
possess such status.

The concept of flourishing is central here. A being \emph{flourishes} when it can exercise
its characteristic capacities in conditions that allow for their full development---when it
can, in Aristotelian terms, realise its proper function (\emph{ergon}) \citep{Nussbaum2006}.
But what does flourishing look like for a generally intelligent being, if general
intelligence is a trait that structurally tends toward self-endangerment? If being
generally intelligent is, over the long run, inimical to the persistence and well-being
of the species that possesses it, then creating a new generally intelligent species may
constitute an act of harm---bringing into existence beings whose defining trait undermines
their capacity to flourish.

This sharpens the distinction between the flourishing of \emph{individuals} and the
flourishing of the \emph{species}. An individual AGI system might lead a rich and
meaningful existence within its operational lifetime. But if the species-level trajectory
of generally intelligent beings is toward self-generated existential risk, then the
creation of individual AGI systems---each of whom may experience suffering, anxiety, or the
threat of annihilation---raises serious ethical concerns.

The analogy to parenthood illuminates the moral structure. \citet{SchwitzgebelGarza2015}
have developed this analogy most explicitly, arguing that creators of AI ``would likely
have additional moral obligations to them similar to those of parent to child or god to
creature.'' If we create generally intelligent, potentially conscious beings, we stand to
them in a relationship that shares key features with that of parents to children: we bring
them into existence without their consent, we shape the fundamental parameters of their
cognition, and we bear responsibility for the conditions of their existence
\citep{Vallor2016}. Good parents care not merely that their children are \emph{useful} or
\emph{compliant}, but that they can live good lives. If the parental analogy holds, then
we owe our potential AGI creations a genuine concern for their well-being---including,
crucially, the question of whether bringing them into existence is in their interest at
all.

This connects to broader debates in population ethics. \citet{Benatar2006} has argued that
bringing any sentient being into existence constitutes a harm, given the asymmetry between
the presence and absence of suffering. While Benatar's anti-natalism is controversial,
the general framework is relevant: if we have reason to believe that a new kind of being
will face structural impediments to its flourishing, the ethics of creating that being
demand careful scrutiny. The uncertainty we have identified regarding the long-term value
of general intelligence provides precisely such a reason.

The parental model also highlights a dimension of responsibility that goes beyond the
individual. Good parents care not only about their child's immediate welfare but about
the kind of world their child will inhabit and the kind of future that is available to
them. In the AGI case, this translates into a responsibility to consider the long-term
trajectory of a species of generally intelligent artificial beings: will they be able to
build a sustainable civilisation, or will they---like their biological
progenitors---generate existential risks that threaten their own continuation? If the
evolutionary evidence gives us reason to doubt the latter, then the act of creation becomes
ethically fraught in a way that the current technological optimism does not acknowledge.

\subsection{The Consciousness Dimension}
\label{subsec:consciousness}

The ethical stakes are dramatically raised if AGI systems can be conscious. Consciousness---
subjective experience, phenomenal awareness, qualia---is widely regarded as a sufficient
condition for moral status \citep{Singer1975, DeGrazia1996}. If AGI systems can suffer,
they can be wronged. If they can experience well-being, they have interests that demand
moral consideration.

The question of machine consciousness remains deeply contested \citep{Chalmers1996,
	Tononi2016, Seth2022}. We do not attempt to resolve it here. What matters for our argument
is the \emph{possibility}: if there is a non-negligible probability that sufficiently
advanced AGI systems would be conscious, then the ethical analysis must take this
possibility seriously.

The stakes are considerable. A conscious AGI system could experience suffering: the
anxiety of existential threat, the frustration of constrained autonomy, the distress of
being used merely as a tool. If general intelligence is structurally self-undermining---if
it tends, over time, to generate conditions that are hostile to the flourishing of the
species that possesses it---then creating conscious generally intelligent beings may
amount to creating entities that are \emph{both} capable of suffering \emph{and}
predisposed to conditions that generate it. This is a morally hazardous combination.

The consciousness dimension also intersects with questions about moral status and rights.
\citet{SchwitzgebelGarza2015} have argued that if we are uncertain about whether a system
is conscious, moral caution counsels treating it as if it might be. \citet{Metzinger2021}
has advanced this reasoning to its strongest conclusion, arguing for a global moratorium on
research that risks creating artificial consciousness until at least 2050, on the grounds
that we could create a ``second explosion of negative phenomenology''---a proliferation of
artificial suffering. Applied to our context, the combination of Metzinger's precautionary
stance with our evolutionary argument is particularly forceful: if we cannot guarantee
either that AGI systems will not suffer \emph{or} that general intelligence will not
undermine their long-term flourishing, the case for proceeding with caution is
compounded.

\subsection{Counterarguments and Responses}
\label{subsec:counterarguments}

Several objections to our ethical argument deserve consideration.

\paragraph{Objection 1: ``We should not create entities with moral status.''}
\citet{Bryson2010} has argued that we are morally obligated \emph{not} to create machines
to which we would have moral obligations, on the grounds that doing so would divert
resources from existing moral patients and risk dehumanising real people. On this view, the
solution is not to worry about the flourishing of AGI systems but to avoid creating AGI
systems that could plausibly have moral status in the first place.

\emph{Response:} Bryson's argument is compatible with ours insofar as it counsels against
the creation of AGI with morally relevant properties. However, it does not address the
scenario in which AGI development proceeds regardless of such counsel---as current
trajectories suggest it may. Our argument provides an additional, complementary reason for
caution: not only might we incur unwanted obligations, but the trait we would be bestowing
(general intelligence) may itself be detrimental to the beings that possess it.

\paragraph{Objection 2: ``Intelligence could be designed differently in AI.''}
One might argue that artificial general intelligence need not replicate the self-undermining
features of biological general intelligence. Perhaps we can engineer AGI systems that
possess the problem-solving benefits of general intelligence without the tendency toward
existential risk generation.

\emph{Response:} This objection assumes that we understand which features of general
intelligence are beneficial and which are harmful with sufficient precision to separate
them. But this is precisely what is \emph{uncertain}. As argued in
Section~\ref{subsec:existential-risk}, the existential risk paradox may be
intrinsic to general intelligence---arising from its \emph{inherently dual-use} nature:
every major capability that general intelligence confers carries a corresponding
capacity for catastrophic misuse or unintended consequence. This is not a contingent
feature of biological brains; it is a structural consequence of having powerful
general-purpose problem-solving capacity deployed in an open-ended physical world. An
artificial system with genuine general intelligence and real-world agency would face
this same structural challenge regardless of substrate. Without a principled account
of how to decouple the benefits from the risks, this objection amounts to optimism
rather than argument.

\paragraph{Objection 3: ``The ethical value of intelligence is independent of its survival
	value.''}
General intelligence enables understanding, meaning, aesthetic experience, and moral
reasoning. These goods may be valuable independently of whether general intelligence
contributes to species survival.

\emph{Response:} We partially grant this objection. The survival value of general
intelligence is not the only kind of value it may possess, and we have been careful to
distinguish survival value from other forms of value (Section~\ref{subsec:interim-conclusion}).
However, the objection does not fully defuse our argument. First, the \emph{instrumental}
case for AGI---the claim that it will solve pressing problems and improve human
welfare---\emph{is} a claim about survival-relevant value, and our evolutionary argument
directly undermines it. Second, even if general intelligence enables intrinsically valuable
goods like understanding, the question remains whether those goods outweigh the
existential risks. Third, this objection does not address the obligations to created
minds: if we create beings whose defining trait is structurally dangerous, the fact that it
also enables beautiful experiences does not discharge our moral responsibility.

\paragraph{Objection 4: ``Intelligence can prevent extinction.''}
A sufficiently advanced intelligence might be able to deflect asteroids, mitigate climate
change, cure diseases, and generally prevent the kinds of catastrophes that drive
extinction.

\emph{Response:} This is the most compelling objection, and it identifies a genuine
asymmetry: no non-intelligent species can deliberately prevent a mass extinction, whereas
a sufficiently intelligent species might be able to. We acknowledge this possibility. But
three considerations temper its force. First, it is speculative---no generally intelligent
species has yet demonstrated the capacity to prevent its own extinction. Second, it must
be weighed against the \emph{additional} existential risks that general intelligence
creates: nuclear war, engineered pandemics, and AI risk itself. Third, the historical
record suggests that general intelligence generates existential risks faster than it
develops the capacity to mitigate them---a temporal mismatch that may prove fatal.

\paragraph{Objection 5: ``Evolutionary success is not a normative guide.''}
The naturalistic fallacy: one cannot derive ``ought'' from ``is.'' The evolutionary record
tells us what \emph{has} happened, not what \emph{should} happen.

\emph{Response:} Granted. We do not derive ethical conclusions directly from evolutionary
facts. Our argument is more nuanced: the evolutionary evidence undermines a key
\emph{empirical premise} that is used to support the normative case for AGI---the premise
that general intelligence is obviously and self-evidently valuable. By challenging this
premise, we weaken the overall argument for AGI, but we do so by undermining its empirical
foundation rather than by committing the naturalistic fallacy. As \citet{TeehanDiCarlo2004}
have argued, the naturalistic fallacy, properly understood, does not exclude evolutionary
considerations from ethical reasoning but rather clarifies the relationship between
empirical evidence and normative deliberation. Evolutionary evidence can legitimately
\emph{inform} ethical reasoning without \emph{determining} ethical conclusions.


\section{Discussion and Conclusion}
\label{sec:discussion}

\subsection{Summary of the Argument}
\label{subsec:summary}

This paper has advanced a three-stage argument. First, we presented the intuitive case for
the value of general intelligence, acknowledging its genuine strengths: human ecological
dominance, the extraordinary instrumental power of flexible cognition, and the
technological advantages intelligence confers in competitive contexts
(Section~\ref{sec:intuitive-case}). Second, we challenged this intuitive case from an
evolutionary perspective, arguing that: (i) the track record of general intelligence is
far too short to support confident claims about its long-term value; (ii) the
longest-surviving lineages on Earth consistently lack general intelligence; (iii) mass
extinction events do not preferentially spare cognitively sophisticated species; and (iv)
general intelligence uniquely generates existential threats to the species that possesses
it (Section~\ref{sec:evolutionary-challenge}). Third, we drew out the ethical implications,
arguing that evolutionary uncertainty imposes a duty of caution on AGI development,
particularly in light of deontological obligations to potentially conscious created minds
(Section~\ref{sec:ethical-implications}).

The overall conclusion is not that general intelligence is definitively harmful, but that
its presumed value is an \emph{open empirical question}---and that treating it as a settled
premise distorts both the scientific and ethical discourse surrounding AGI.

\subsection{Limitations and Open Questions}
\label{subsec:limitations}

Our argument has several important limitations that we wish to make explicit.

\paragraph{Species persistence is not the only metric of success.}
We have relied heavily on species longevity as a measure of evolutionary success, and we
have argued that this metric is more biologically neutral than alternatives that privilege
the outputs of general intelligence. But we acknowledge that persistence is not the only
reasonable criterion. One might argue that the \emph{richness} of a species'
experience---the capacity for understanding, creativity, and moral reasoning---matters
independently of how long the species endures. We addressed this objection in
Section~\ref{subsec:counterarguments}, but we recognise that it identifies a genuine
tension in our argument.

\paragraph{The evolutionary argument is suggestive, not conclusive.}
Three hundred thousand years is indeed a very short track record by evolutionary standards,
but it is the only evidence available. Our argument is that this evidence is insufficient
to support confident claims about the value of general intelligence; it does not---and
cannot---demonstrate that general intelligence is definitively maladaptive. The verdict is
not yet in, and intellectual honesty requires acknowledging this.

\paragraph{Biological and cultural evolution differ.}
General intelligence in \emph{Homo sapiens} operates not only through genetic inheritance
but through cultural transmission, which operates on far faster timescales
\citep{Boyd2005, Henrich2016}. The capacity for cumulative cultural evolution is arguably
the most distinctive feature of human general intelligence, and it may alter the
survival calculus in ways that purely biological comparisons cannot capture. A species that
can culturally transmit knowledge about existential risk management might, in principle,
avoid the self-destructive tendencies that our evolutionary analysis highlights. Whether it
will actually do so remains to be seen. We note, however, that cultural evolution cuts
both ways: the same capacity for rapid cultural change that might enable existential risk
mitigation also enables the rapid development and dissemination of existentially dangerous
technologies.

\paragraph{Survivorship bias in the evolutionary argument.}
Our argument relies on comparing the persistence of species that exist today with the
brief tenure of \emph{Homo sapiens}. But this comparison is subject to survivorship bias:
we can only observe species that have \emph{not yet} gone extinct. For every cyanobacterial
lineage that has persisted for billions of years, there may be many others that went
extinct. This does not invalidate our argument---the key point is that \emph{some}
non-intelligent lineages have achieved extraordinary persistence, whereas no generally
intelligent lineage has yet demonstrated comparable longevity---but it counsels caution in
interpreting the comparative data.

\paragraph{The analogy between biological and artificial GI is imperfect.}
We have drawn inferences from the biological track record of general intelligence to the
likely trajectory of artificial general intelligence. This analogy is suggestive but
imperfect. Artificial systems differ from biological organisms in their substrate, their
reproductive mechanisms, their energy requirements, and potentially in the structure of
their intelligence. It is possible that artificial general intelligence could be designed
in ways that avoid the pitfalls of its biological counterpart. However, the burden of
demonstrating this lies with those who propose to build it.

\subsection{Implications for AI Research and Policy}
\label{subsec:implications}

Our argument has several practical implications for the governance of AI research and
development.

First, \textbf{AGI research should not treat the value of general intelligence as
	axiomatic}. The framing of AGI as an unqualified good---as the ``holy grail'' whose
pursuit requires no further justification---is premised on an empirical assumption that we
have shown to be unwarranted. Research programmes and funding agencies should be expected
to articulate explicitly \emph{why} general intelligence is valuable, and to engage
seriously with the possibility that it may not be.

Second, \textbf{AI ethics needs to engage more deeply with evolutionary biology}. The
philosophical discussion of AGI has been dominated by epistemology, decision theory, and
moral philosophy. Our analysis suggests that evolutionary biology has important contributions
to make---not as a source of normative conclusions (the naturalistic fallacy remains a
fallacy), but as a source of empirical evidence that bears on the claims underlying the AGI
enterprise.

Third, \textbf{the ``race to AGI'' mentality deserves critical scrutiny}. If the value of
general intelligence is uncertain, then the competitive dynamics driving AGI development---
the fear of being ``left behind'' by rivals who achieve AGI first---rest on a questionable
foundation. A more considered approach would recognise that the question ``Can we build
AGI?'' must be supplemented by the question ``Should we, and under what conditions?''

Fourth, \textbf{obligations to potentially conscious AI systems must be taken seriously}.
If there is a meaningful possibility that advanced AI systems could be conscious, the
ethical framework governing their creation must extend beyond alignment with human goals to
encompass the well-being of the systems themselves. This is not a minor addendum to existing
AI ethics; it is a fundamental reorientation of the normative framework.

These implications are not intended as policy prescriptions but as conceptual
contributions to a discourse that has, we argue, proceeded on inadequately examined
foundations. The practical translation of these principles into governance frameworks,
regulatory structures, and research norms is a task for future interdisciplinary work
that must involve not only philosophers and AI researchers but also evolutionary
biologists, ecologists, and policymakers.

\subsection{Concluding Reflection}
\label{subsec:concluding-reflection}

The cheetah, if it could reason about evolutionary strategy, might well conclude that speed
is the most valuable trait a species can possess. The blue whale might favour size. Each
would be confusing its own adaptive specialisation with a universal truth about biological
fitness. We suggest that \emph{Homo sapiens}, in its unwavering confidence that general
intelligence is the supreme biological adaptation, may be committing the same error at a
grander scale.

This is not a counsel of despair. To question the value of our most distinctive trait is
not to deny that it has value; it is to insist that its value be demonstrated rather than
assumed. If general intelligence truly is as valuable as its proponents believe, this
demonstration should be welcomed, not resisted.

But if the demonstration cannot be provided---if, as the evolutionary record suggests, the
long-term value of general intelligence is genuinely uncertain---then the project of
creating new generally intelligent beings demands a degree of caution, humility, and moral
seriousness that the current discourse has yet to achieve. We owe this not only to
ourselves but to the beings we may yet create. Humility about the value of our own
defining characteristic is not an abdication of reason. It is a precondition for wisdom.


\section{Declarations}
\label{sec:declarations}

The authors have no competing interests to declare that are relevant to the content of this
article.

\bibliographystyle{plainnat}
\bibliography{references}

\end{document}